\documentstyle[prl,tighten,aps,epsf,graphics,multicol,amstex]{revtex}
\begin{document}

\draft

\title{Classical Information and Distillable Entanglement}
\author{J. Eisert$^{1}$, T. Felbinger$^{1}$, P. Papadopoulos$^{2}$, 
M.B. Plenio$^{2}$, and M. Wilkens$^{1}$}
\address{${(1)}$
Institut f{\"u}r Physik, Universit{\"a}t Potsdam, 
14469 Potsdam, Germany \\
${(2)}$ Optics Section, Blackett Laboratory, Imperial
College, London SW7 2BZ, United Kingdom}

\date{\today}
\maketitle

\begin{abstract}
We establish a quantitative connection between the amount of lost
classical information about a quantum state and concomitant loss 
of entanglement. Using methods that have been developed for the 
optimal purification of mixed states we find a class of mixed states 
with known distillable entanglement. These results can be used to determine
the quantum capacity of a quantum channel which randomizes the order of
transmitted signals. 
\end{abstract}


\begin{multicols}{2}
\narrowtext

The development of quantum information processing in recent 
years has shown that quantum information and in particular
quantum entanglement allow for the realization of applications
that are not possible classically \cite{Reviews}. 
Classical information has however not become obsolete as a simple 
limiting case of the more general theory. In fact, there 
are interesting connections for example between the amount 
of quantum entanglement \cite{Quant,Quant1,QuantLong}
that is held by two parties and the classical information 
that is available about the jointly held system \cite{Cohen}. 
An extreme example would be one where the two parties Alice 
and Bob are sharing an equal mixture of two Bell states. 
Being completely ignorant about the identity of the 
state, the density operator describing the system can also
be described as an equal mixture of two product states, which
implies that Alice and Bob share no entanglement at all. However, 
given the one bit of information about the identity of the state, 
they share one ebit of entanglement (one maximally entangled 
state between two qubits). While the exact relation between the 
amount of classical information required per gained ebit is 
unknown (see also \cite{Cohen}), this example illustrates that 
the retrieval of classical information can lead to an increase 
in the usable entanglement. Quite analogously the loss of 
classical information will usually reduce the amount of 
entanglement held between two parties. In this paper we will 
consider a particularly clear way in which classical information 
is lost. Surprisingly, for the resulting class of mixed states
the distillable entanglement can be determined. This example can 
also be interpreted as a noisy quantum channel which randomizes 
the order of transmitted signals. Using the results of this
paper, we are able to determine the quantum capacity of such a 
quantum channel. 

Imagine two spatially separated parties, Alice and Bob, who are
holding two entangled pairs of particles which they would like 
to use later on, for example to implement some quantum communication
protocol. As Alice and Bob share two pairs of identical particles, 
they need a classical record about the order of the particles. This 
means that it is known which of Alices particles is entangled with 
which particle of Bob (see left part of Fig. 1 which represents the 
mixed state of apparatus and system where the ancilla 
allows to determine the order of the particles). Now imagine that by 
some misfortune (e.g. the particles in a transmission arrive in random order), 
this classical record is destroyed, i.e., the ancilla is unavailable! 
In that case the state of the two pairs kept by Alice and Bob is an 
equal mixture of two possible states: One where the first of 
Alice's particles is entangled with the first of Bob's particles and 
another where the first of Alice's particles is entangled with the 
second of Bob's particles. This implies that Alice and Bob have less 
information about each other's systems, i.e., that the mutual information 
is reduced. In the context of a quantum channel one would have a situation 
where signals change their order randomly. The natural question for Alice 
and Bob is then whether they are still holding quantum mechanical 
entanglement and if yes how much, i.e., what is the capacity of the
associated quantum channel. Let us state the issue more 
formally.

{\it When information $\Delta I$ about the order of a number of 
entangled quantum systems is lost, is the resulting state of any 
use for quantum communication purposes? How much 
entanglement $\Delta E=E_{\rm{before}}-E_{\rm{after}}$ 
has been destroyed and what can be said about the ratio $\Delta E/\Delta I$?}

It is this question that we will investigate in this work 
to explore the connection between classical and quantum 
information. First we will answer the special case in which 
Alice and Bob are holding two pairs of maximally entangled
states. Subsequently we will solve the case of an even
number of copies of pairs of arbitrarily entangled particles. 
These results then give rise to a bound for the ratio 
$\Delta E/\Delta I$ in a more general situation. 
The concept of entanglement which is employed in the following 
is that of distillable entanglement $E_D$ \cite{Quant,QuantLong,Rains,Bennett} 
with respect to separable operations \cite{Kraus}. This means that 
we are interested in the maximal rate with which entanglement purification
can obtain maximally entangled states from a state which has arisen due to
the loss of classical information. \\

\noindent {\bf Example. --} 
Consider the situation where Alice and Bob share two pairs 
of two-level systems -- i.e., qubits -- each in a maximally 
entangled state of the form $(|00\rangle + |11\rangle)/\sqrt{2}$. 
In this way they are sharing two ebits of entanglement. Now Bob 
loses the information about the order of his quantum systems. 
This means that Bob does not know whether his two particles are 
in the original order or have been permuted (see right half of Fig.\ 1). 
\begin{figure}
\centerline{
        \epsfxsize=6.5cm
       \epsfbox{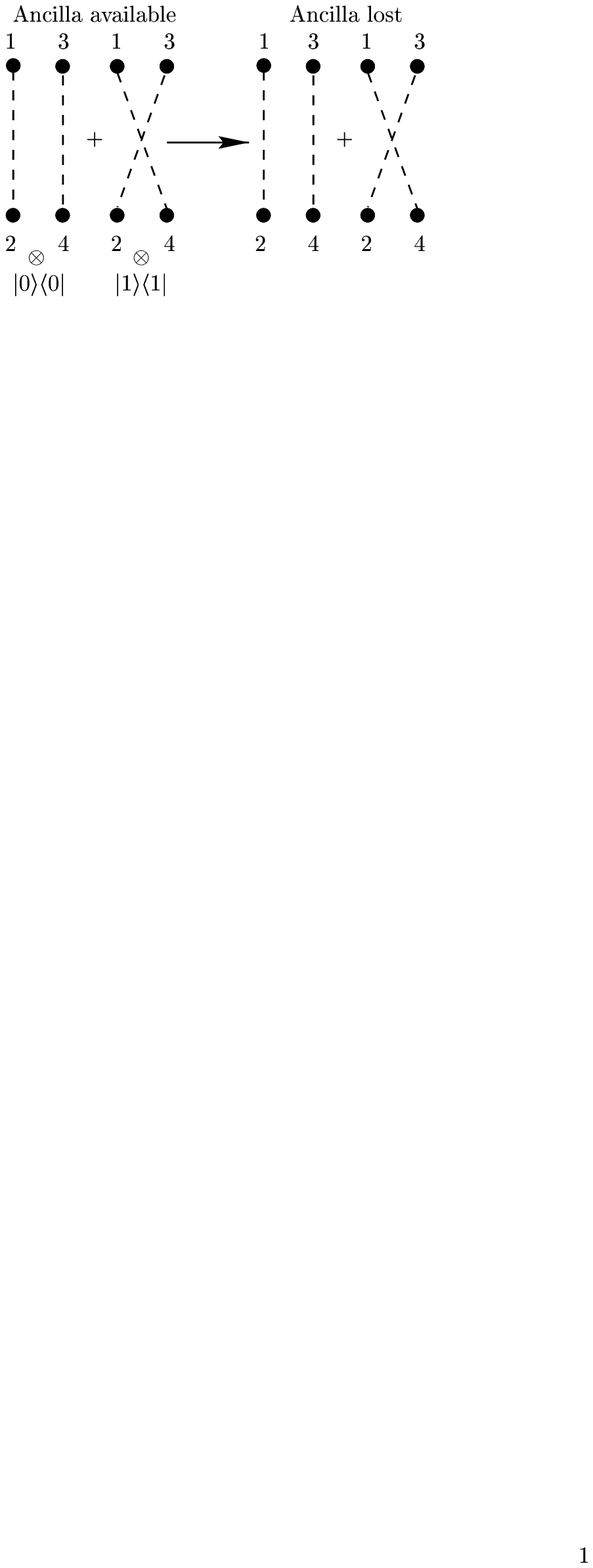}
}
\vspace*{0.2cm}
\caption{In the left half of the figure system and ancilla are in 
a mixed state. The ancilla allows to determine the order of the 
particles -- with 
probability $p=\frac{1}{2}$ the first particle of Alice either 
is entangled with the first of Bob or with the second 
particle of Bob. On the right hand side, the ancilla 
is lost and one cannot determine the order anymore.
The information that the ancilla had about the system is quantified
by the mutual information between system and ancilla. }
\label{fig:Scheme}
\end{figure}

Let $|\psi_1\rangle\langle\psi_1|$ be the state of the qubits 
labelled $1$, $2$, $3$, and $4$ in the original situation (see Fig.\ 1). 
$1$ and $3$ are Alice's qubits, Bob's qubits are numbered $2$ and $4$. 
In the computational basis $|\psi_1\rangle$ is given by
\begin{eqnarray}
	|\psi_1\rangle=(|0000\rangle+
	|0011\rangle+
	|1100\rangle+
	|1111\rangle)/2 \;\; ,
\end{eqnarray}
while in the permuted case where the role of $2$ and $4$
is interchanged the state is
\begin{eqnarray}
	|\psi_2\rangle=
	(|0000\rangle+
	|0110\rangle+
	|1001\rangle+
	|1111\rangle)/2 \;\; .
\end{eqnarray}
As a result of the loss of the order of the particles
on Bob's side, the composite quantum system is now described by the
density operator 
\begin{equation}\label{sigma}
	\sigma=(|\psi_1\rangle\langle\psi_1|+|\psi_2\rangle\langle\psi_2|)/2
	\;\; .
\end{equation}
It is now natural to ask how much entanglement is still accessible
to Alice and Bob, i.e., how much distillable entanglement the
state $\sigma$ holds. To solve this question consider the spectral 
decomposition of $\sigma$ given by
\begin{equation}
	\sigma= \frac{1}{4}|\phi_1\rangle\langle\phi_1| +
	\frac{3}{4}	|\phi_2\rangle\langle\phi_2| \;\; ,
\end{equation}
where 
\begin{mathletters}
\begin{eqnarray}
	|\phi_1\rangle&=&
	(|0011\rangle-
	|0110\rangle-
	|1001\rangle+
	|1100\rangle
	)/2,\\
	|\phi_2\rangle&=&
	(2|0000\rangle+
	|0011\rangle+
	|0110\rangle\nonumber\\
	&&+
	|1001\rangle+
	|1100\rangle+
	2|1111\rangle)/\sqrt{12} \;\; .
\end{eqnarray}
In the basis of angular momentum eigenstates $|j,m\rangle$ 
with $j=0,1$; $m=-1,0,1$ which is given by
\end{mathletters}
\begin{mathletters}
\begin{eqnarray}
	|1,-1\rangle=|00\rangle,\,\,\,\,&&
	|1,1\rangle=|11\rangle,\\
	|1,0\rangle=(|01\rangle+|10\rangle)/\sqrt{2},\,\,\,\,&&
	|0,0\rangle=(|01\rangle-|10\rangle)/\sqrt{2} \;\; ,
\end{eqnarray}
the eigenstates $|\phi_1\rangle$ and $|\phi_2\rangle$ read
\end{mathletters}
\begin{mathletters}
\begin{eqnarray}
	|\phi_1\rangle&=&
	|0,0\rangle|0,0\rangle,\\
	|\phi_2\rangle&=&
	\left(
	|1,-1\rangle|1,-1\rangle
	+
	|1,0\rangle|1,0\rangle
	+
	|1,1\rangle|1,1\rangle
	\right)/\sqrt{3} \;\; .
\end{eqnarray}
\end{mathletters}
Here, the first ket corresponds to Alice's qubits, 
the second ket to Bob's.

An upper bound \cite{QuantLong,Rains} for the distillable 
entanglement is given by the relative entropy of entanglement 
\cite{Quant,QuantLong} $E_R(\sigma)$ of $\sigma$ which in turn 
is smaller or equal to the relative entropy with respect
to any separable state $\rho$. Hence, the distillable
entanglement $E_D(\sigma)$ of $\sigma$ is bounded by
\begin{equation}
	E_D(\sigma) \le S(\sigma||\rho)= \frac{3}{4}\log 3,
	\label{bound1}
\end{equation}
where the disentangled state $\rho$ is chosen as
$
	\rho = \frac{1}{4} \sum_{j=0}^{1} \sum_{m=-j}^{j} |j,m\rangle|j,m\rangle\langle j,m|\langle j,m| \, .
$
Surprisingly, it turns out 
that the upper bound given in Eq. (\ref{bound1}) can indeed be 
achieved! 

In the optimal distillation protocol 
Alice performs a von-Neumann projective measurement
with the two possible projectors $A_1=|0,0\rangle\langle0,0|$ and 
$A_2=	\sum_{m=-1}^{1} |j=1,m\rangle\langle j=1,m|$,
while Bob remains inactive, i.e., $B_1=B_2={\Bbb{I}}_B$. 
With probability $p_1=1/4$ they obtain the normalized 
output state $|\phi_1\rangle\langle\phi_1|$, which is a product
state and of no further use. With probability $p_2=3/4$ they 
obtain $|\phi_2\rangle\langle\phi_2|$ which has $\log 3$ ebits 
of entanglement. The average number of maximally entangled 
states that can be distilled from $\sigma$ is given by 
\begin{equation}
 	E_D(\sigma)=(3/4)\log3 \approx 1.189\;\; .
\end{equation}
As this realizes the bound Eq. (\ref{bound1}) it is the maximally 
possible value \cite{reason}. It is worth noting that this value 
is greater than one. Hence, less than one ebit of entanglement is 
erased due to the loss of the classical information about 
the order. The classical information can be calculated as the mutual
information between the ancilla and the system of Alice and Bob. 
This turns out to be equal to the entropy of the mixed state that 
Alice and Bob share afterwards (i.e., $\Delta I = S(\sigma)$). Therefore 
we find
\begin{equation}
	\frac{\Delta E_D}{\Delta I} = 1 \;\; . \label{equal}
\end{equation}
The above scenario can be generalized to the situation where Alice 
and Bob initially do not hold maximally entangled states but pure 
states of the type $\alpha|00\rangle+\beta|11\rangle$ \cite{Schmidt}
with a given degree of entanglement. This case is interesting 
since it leads to an operationally defined one parameter 
class of states for which the distillable entanglement with 
respect to separable operations can be analytically computed.
Here, we consider
$
\sigma=(|\psi_1\rangle\langle\psi_1|+|\psi_2\rangle\langle\psi_2|)/2
$
with
\begin{eqnarray}
	|\psi_1\rangle&=&\alpha^2 |0000\rangle+
	\alpha\beta |0011\rangle+
	\alpha\beta|1100\rangle+
	\beta^2|1111\rangle,\\
	|\psi_2\rangle&=&
	\alpha^2|0000\rangle+
	\alpha\beta|0110\rangle+
	\alpha\beta|1001\rangle+
	\beta^2|1111\rangle,
\end{eqnarray}
where $\alpha\in[0,1]$, $\beta=\sqrt{1-\alpha^2}$.
Following the previous calculation, we find that the distillable 
entanglement is 
\begin{eqnarray}
	E_D(\sigma)=E_R(\sigma)&=&
	(1-\alpha^2 \beta^2)
	\log( 1-\alpha^2 \beta^2)\label{Full}\\
	&-&
	(\alpha^4\log\alpha^4+
	\beta^4 \log \beta^4
	+
	\alpha^2 \beta^2 \log \alpha^2 \beta^2).\nonumber
\end{eqnarray}
Since the entanglement of the initial pure state was given by the 
entropy of the reduced states of Alice or Bob, that is, by 
$-2(\alpha^2 \log \alpha^2+ \beta^2 \log \beta^2)$, again, 
$\Delta E_D/\Delta I=1$ for all $\alpha\in[0,1]$. Of course, 
Eq.\ (\ref{Full}) reduces to $E_D(\sigma)= (3/4)\log 3$ for
$\alpha=\beta=1/\sqrt{2}$.

The example we have presented here leads to a more general
proposition which we are going to prove here. We restrict
the argument to the case where Alice and Bob are initially 
sharing an even number of pairs in pure states with 
two-particle entanglement between qubits.\\

\noindent {\bf Proposition. --} Let Alice and Bob share 
$N=2J$ pairs of qubits each in the same state $|\phi\rangle$.
The associated Hilbert space is ${\cal H}={\cal H}_A\otimes {\cal H}_B\sim
({\Bbb{C}}^2)^{\otimes 2J}\otimes ({\Bbb{C}}^2)^{\otimes 2J}$,
$J=1,2,...$. Bob then loses the information about the order of the 
qubits completely, the information loss $\Delta I$ being quantified by the mutual 
information between ancilla and system. As a consequence, 
Alice and Bob are now sharing a less entangled mixed state $\sigma$. 
The distillable entanglement of the state $\sigma$ can be calculated
exactly and the ratio between the change of distillable $\Delta E_D$ and the
amount of erased information $\Delta I$ obeys for any $J=1,2,...$ the inequality
\begin{equation}\label{ineq}
	\frac{\Delta E_D}{\Delta I}\le 1 \;\; ,
	\label{eqprop}
\end{equation}
with equality for $J=1$.
 
{\it Proof:} Let us first consider the case where $|\phi\rangle$ is 
a maximally entangled state $|\phi\rangle = (|00\rangle+|11\rangle)/\sqrt{2}$. 
To prove the statement of the proposition 
we first construct the state $\sigma$ after the loss of the order of 
Bob's particles. Then the optimal entanglement purification protocol 
will be presented and its optimality proven. The loss in information
can be calculated and the validity of Eq.\ (\ref{ineq}) for these 
particular initial states is then validated. In the more general case
of arbitrary initial states $|\phi\rangle = \alpha|00\rangle +\beta|11\rangle$ 
the same approach can be applied, confirming Eq. (\ref{eqprop}).

Let ${\cal H}={\cal H}_A \otimes {\cal H}_B$ be the underlying Hilbert space 
and ${\cal S}( {\cal H})$ the associated state space. Since ${\cal H}_A$ and 
${\cal H}_B$ are $2J$-fold tensor products of Hilbert spaces isomorphic to 
${\Bbb{C}}^2$, they can be decomposed into a direct sum of orthogonal
subspaces of the form 
$
	{\cal H}_A= \bigoplus_{j=0}^J
	\bigoplus_{\alpha_{j}}{\cal H}^A_{j,\alpha_{j}},\,\,\,
	{\cal H}_B=
	\bigoplus_{k=0}^J
	\bigoplus_{\beta_{k}}{\cal H}^B_{k,\beta_{k}},
$
where 
${\cal H}^A_{j,\alpha_j}=\text{span}
	\{|j,m,\alpha_j\rangle| m=-j,-j+1,...,j\}$ 
	for $j=0,1,...,J$ and $\alpha_j=1,2,...,d_j$.
The additional degeneracy is 
given by 
$
	d_j = \frac{2j+1}{2J+1} \left( {2J+1 \atop J-j } \right) \;\; .
$
The actual representation is clearly dependent on the choice of basis.
As in \cite{Cir} we choose $|j,m,1\rangle=|j,m\rangle\otimes 
[(|01\rangle-|10\rangle)/\sqrt{2}]^{\otimes(J-j)}$,
where $|j,m\rangle$ is the state of $2j$ qubits with a fixed value of 
$j$ and $m$ with $j-m$ qubits in $|0\rangle$. ${\cal H}_B$ can be 
decomposed into a direct sum in exactly the same fashion. 

Using this decomposition of the Hilbert space, it can easily be
seen that the initial state of the $N$ pairs shared between Alice
and Bob can be written as
\begin{equation}
	\otimes_{n=1}^N |\phi\rangle 
	= \sum_{j,m,\alpha }^{} \sqrt{\frac{1}{2^{2J}}}
	|j,m,\alpha\rangle |j,m,\alpha\rangle \;\; .
	\label{initialstate}
\end{equation}
The state $\sigma$ after the loss of the order of Bobs particles
$\sigma$ is then given by 
$
	\sigma=\sum_{j=0}^J\sum_{\alpha_j, \beta_j=1}^{d_j} 
	p_j
	|\psi_j (\alpha_j, \beta_j)\rangle
	\langle \psi_j (\alpha_j, \beta_j)|
$, with
\begin{equation}
	|\psi_j (\alpha_j, \beta_j)\rangle=
	\frac{1}{\sqrt{2j+1}}\sum_{m=-j}^j 
	|j,m,\alpha_j\rangle
	|j,m,\beta_j\rangle
\end{equation}
and $p_j=(2j+1)/(2^{2J} d_j)$.
This particular form of the state after loss of the order of the particles
can be proven using Schur's first lemma \cite{Messiah}.

As before, the following distillation protocol is based on
the fact that the subspaces of the state space corresponding
to the above components of the underlying Hilbert space
are locally distinguishable. Interesting enough, this
protocol is related to the algorithm proposed in \cite{Cir}
for the optimal purification of qubits.

1.) Alice performs a local projection
measurement in such a way
that her reduced state is element of
${\cal S}({\cal H}^A_{j,\alpha_j})$ for
some $j=0,...,J$; $\alpha_j=1,...,d_j$.

2.) If $\alpha_j\neq1$ she applies a local unitary operation 
$U^A_{j,\alpha_j}$ such that her reduced state is included
in the set ${\cal S}({\cal H}^A_{j,1})$. Since in general
$|j,m,\alpha_j\rangle$ is a linear superposition 
of $\Pi_i|j,m,1\rangle$, where $\Pi_i$, $i=1,2,...$ are
appropriate locally acting
permutation operators, this is always possible.

3.) The reduced state $\sigma_A$ of Alice is at this stage
of the structure $\sigma_A=\omega_A\otimes 
[(|01\rangle-|10\rangle)(\langle01|-\langle 10|)/2]^{\otimes(J-j)}$.
The last $J-j$ pairs of qubits in the singlet state are neither entangled
with the other qubits on her side nor entangled with any of Bob's qubits.
Hence, they will be of no further use in the distillation protocol.

4.) Bob performs a local measurement projecting his reduced state on 
${\cal S}({\cal H}^B_{k,\beta_k})$ with some $k=0,1,...,J$ and 
$\beta_k=1,2,...,d_k$. Due to the particular form of the initial state
$k=j$, but he may get a $\beta_j$ different from the $\alpha_j$ obtained
by Alice.

5.) In the same way as before Bob applies a 
local unitary operation $U^B_{j,\beta_j}$ such that his
reduced state is element of ${\cal S}({\cal H}^B_{j,1})$.

6.) Alice and Bob end up with the probability
$d_j^2 p_j=(2j+1) d_j/(2^{2J})$ in one of the pure states 
$|\psi_j\rangle\langle\psi_j|$, where
\begin{equation}
|\psi_j\rangle = \frac{1}{\sqrt{2j+1}}\sum_{m=-j}^j |j,m,1\rangle|j,m,1\rangle.
\end{equation} 
This state contains $\log (2j+1)$ ebits of entanglement. Hence, the
total average number is $\sum_j d_j^2 p_j S(\text{tr}_A(|\psi_j\rangle\langle\psi_j|))=
\sum_j d_j^2 p_j \log(2j+1)$.

To show that the above protocol is actually optimal we consider the 
relative entropy functional of the state $\sigma$ after permutation
with respect to an appropriate separable state $\rho$. 
The separable state $\rho$ is taken to be 
$
	\rho=
	\sum_j p_j \rho_j , 
$
where 
\begin{equation}
\rho_j=
	\sum_{\alpha_j,\beta_j=1}^{d_j}
	\sum_{m=-j}^j
	\frac{
	(|j,m,\alpha_j\rangle\langle j,m,\alpha_j|\otimes
	|j,m,\beta_j\rangle\langle j,m,\beta_j|)
	}
	{2j+1}.
\end{equation}
Since all subspaces associated with different values of $j$, $m$,
$\alpha_j$, and $\beta_j$ are orthogonal and with 
$
p_j=\frac{2j+1}{d_j 2^{2J}}
$
this expression is given by 
\begin{equation}\label{edfull}
	S(\sigma||\rho)=
	\sum_{j=0}^J d_j^2 p_j \log(2j+1),
\end{equation}
which is identical to 
the value given for the average number of maximally
entangled states obtained when employing the above
procedure, and therefore, also identical to the
distillable entanglement $E_D(\sigma)$ with respect to separable 
operations. Since again the mutual information between system and
ancilla turns out to be $\Delta I= S(\sigma)$,
it follows that $\Delta E_D/\Delta I\leq1$ for all $N$ for this 
particular initial state. 

For any pure state we can set
$|\phi\rangle=\alpha|00\rangle+\beta|11\rangle$
with $\alpha\in[0,1]$, $\beta=\sqrt{1-\alpha^2}$ \cite{Schmidt}. 
The same argument as before holds, and the
same protocol is optimal for the distillation with
respect to separable operations. 
The state $\sigma$ after permutation is found to be
$
	\sigma=\sum_{j=0}^J p_j 
	|\psi_j(\alpha_j,\beta_j)\rangle
	\langle  \psi_j(\alpha_j,\beta_j)|,
$
where now $p_j=\sum_{m=-j}^j \alpha^{2(J-m)}\beta^{2(J+m)}/d_j$;
the (unnormalized) states 
$|\psi_j(\alpha_j,\beta_j)\rangle\langle \psi_j(\alpha_j,\beta_j)|$
are defined as $|\psi_j(\alpha_j,\beta_j)\rangle=\sum_{m=-j}^j
	\alpha^{j-m}\beta^{j+m}
	|j,m,\alpha_j\rangle |j,m,\beta_j\rangle$.
Again,
$E_D(\sigma)=
\sum_{j=0}^J d_j^2 p_j S(\text{tr}_A
	|\psi_j(1,1)\rangle \langle \psi_j(1,1)|)$
and $\Delta I=S(\sigma)$. From this and the fact that initially, 
$E_D(|\phi\rangle\langle\phi|^N)=
-N(\alpha^2 \log \alpha^2+ \beta^2 \log \beta^2)$
ebits of entanglement are present, it follows that
also in this case $\Delta E_D/\Delta I\leq 1$ for all $N$ ${}_{\Box}$.

Hence, in scenarios of the type discussed in the proposition,
concomitant with the loss of a number of bits of classical 
information, not more than one ebit of distillable entanglement
is destroyed per bit of classical information.
These findings (see also \cite{Cohen}) lead us to the following\\

\noindent {\bf Conjecture. --} Let $\omega$ be a state 
of a bipartite quantum system taken from a set of (possibly entangled)
states $\omega_1,\ldots,\omega_N$, each of which is assigned 
a classical probability $p_1,..., p_N$. After the loss of the classical 
information about the identity of the state $\omega$ the state of 
the quantum system is taken to be $\sigma=\sum_{n=1}^N p_n \omega_n$.
The change in distillable entanglement 
$\Delta E_D=E_D(\omega)-E_D(\sigma)$ and the loss of
classical information $\Delta I=S(\sigma)$ then obey the inequality 
$\Delta E_D/\Delta I\leq 1$.

In summary we have investigated a practically relevant situation in which 
classical information about the order of particles can be lost, e.g. during 
the transmission via a quantum channel. Surprisingly, the general class of 
mixed states obtained from this procedure have known distillable entanglement 
and therefore the corresponding quantum channel has known quantum capacity 
\cite{Nielsen}. 
It turns out that the ratio between the loss of entanglement and the amount of 
classical information lost in such a situation can be related to an inequality
and we conjecture the general validity of this inequality. These results shed 
new light on the relationship between entanglement purification 
and channel capacity on the one hand and classical information on the other.

This work was supported by the DFG, the EPSRC, the European 
TMR Research Networks ERBFMRXCT960066 and ERBFMRXCT960087, 
The Leverhulme Trust, the State Scholarships Foundation of Greece,
and the European Science Foundation. 
We would also like to thank the participants of the A2
Consortial Meeting for fruitful discussions.

\end{multicols}


\begin{thebibliography}{99}
\bibitem{reason} Note that Alice and Bob do not need to communicate at all
in this protocol. The reason can be found in the local
distinguishability of the states $|\phi_1\rangle\langle \phi_1|$ 
and $|\phi_2\rangle\langle \phi_2|$.
\bibitem{Schmidt}
	The Schmidt decomposition of
	a pure state can always be written in this form,
	where the Schmidt coefficients $\alpha$ and
	$\beta$ can be made real and non-negative.
\end{thebibliography}
\end{document}